\documentclass[twocolumn]{el-author}
\usepackage{tabu}
\usepackage{multirow}
\usepackage{caption}
\usepackage{amsmath}
\usepackage{layouts}
\usepackage{amsfonts}
\usepackage{gensymb}
\usepackage{multirow}
\usepackage{subfig}
\usepackage{algorithmic}
\usepackage{algorithm,tabu}

\newcommand{\ua}{\uparrow}
\newcommand{\nc}{\newcommand}
\nc{\da}{\downarrow} \nc{\hc}{\hat{c}} \nc{\hS}{\hat{S}}
\nc{\bra}{\langle} \nc{\ket}{\rangle} \nc{\eq}{equation (\ref}
\nc{\h}{\hat} \nc{\hT}{\h{T}}\nc{\be}{\begin{eqnarray}}
\nc{\ee}{\end{eqnarray}}\nc{\rd}{\textrm{d}}\nc{\e}{eqnarray}\nc{\hR}{\hat{R}}\nc{\Tr}{\mathrm{Tr}}
\nc{\tS}{\tilde{S}}\nc{\tr}{\mathrm{tr}}\nc{\8}{\infty}\nc{\lgs}{\bra\ua,\phi|}\nc{\rgs}{|\ua,\phi\ket}
\nc{\hU}{\hat{U}}\nc{\lfs}{\bra\phi|}\nc{\rfs}{|\phi\ket}\nc{\hZ}{\hat{Z}}\nc{\hd}{\hat{d}}\nc{\mD}{\mathcal{D}}
\nc{\bd}{\bar{d}}\nc{\bc}{\bar{c}}\nc{\mc}{\mathcal}\nc{\ea}{eqnarray}\nc{\mG}{\mathcal{G}}\nc{\bce}{\begin{center}}
\nc{\ece}{\end{center}}

\begin{document}

\title{BitMix: Data Augmentation for Image Steganalysis}

\author{I.-J. Yu, W. Ahn, S.-H. Nam, and H.-K. Lee}

\abstract{
Convolutional neural networks (CNN) for image steganalysis demonstrate better performances with employing concepts from high-level vision tasks. 
The major employed concept is to use data augmentation to avoid overfitting due to limited data.
To augment data without damaging the message embedding, only rotating multiples of 90\degree or horizontally flipping are used in steganalysis, which generates eight fixed results from one sample.
To overcome this limitation, we propose \textbf{BitMix}, a data augmentation method for spatial image steganalysis.
BitMix mixes a cover and stego image pair by swapping the random patch and generates an embedding adaptive label with the ratio of the number of pixels modified in the swapped patch to those in the cover-stego pair. 
We explore optimal hyperparameters, the ratio of applying BitMix in the mini-batch, and the size of the bounding box for swapping patch.
The results reveal that using BitMix improves the performance of spatial image steganalysis and better than other data augmentation methods.
}
\maketitle

\section{Introduction}
Image steganalysis has been developed to detect hidden messages from an image embedded with steganography (stego), while steganography has been developed to hide messages in the pixels of the textual regions rather than flat regions to minimize the detectability of embedding to the original image (cover).
To detect such a low-level signal, steganalysis uses machine learning, such as convolutional neural networks (CNN).
Xu \emph{et al.} proposed effective architecture with competitive performance \cite{xu2016structural}.
To boost performance, \cite{xu2016ensemble} devised an ensemble methods of multiple detectors with the same architecture.
Meanwhile, CNNs are easy to overfit to a training set, and  many methods can be employed to improve generalizability.
The representative example is data augmentation (DA), which uses signal processing that tweaks samples by shearing, blurring, and adding noise to increase the number of samples for high-level vision tasks.
However, because the traces of steganography are too subtle compared to the image content, the aforementioned DA methods manipulate the stego image more strongly than the steganography itself.
Training this sample as a stego class misguides the network and degrades the performance of steganalysis, which is why there has been very limited use of DA methods in steganalysis.

The current state-of-the-art methods in CNN-based image steganalysis are SRNet \cite{Boroumand2019} and ZhuNet \cite{Zhang2020}, where SRNet is an end-to-end learning network consisting of multiple unpooled layers to focus on low-level features, and ZhuNet preprocesses images with SRM filters and uses depth-wise separable convolutions.
These methods improved the performance with the exquisite network architecture, but they augmented an image to eight different samples that are pixel-preserved by horizontally flipping or rotating multiples of 90\degree, which was proposed in \cite{ye2017deep, yedroudj2018yedroudj}.
Although they reported that simple augmentation facilitates performance improvement, it generates a fixed number of images, which is insufficient to prevent overfitting.

On the other hand, several DA methods with regional dropouts or using two different samples with soft-labels \cite{zhang2017mixup,yun2019cutmix} have been proposed for high-level vision tasks.
MixUp \cite{zhang2017mixup} interpolates two different images, and CutMix \cite{yun2019cutmix} swaps random patches between different images. 
They generate mixed images and soft labels with a ratio of interpolation and swap area.
In terms of DA for low-level vision tasks, CutBlur \cite{yoo2020rethinking} is a representative work that has been proposed for image super-resolution tasks.
It blends low- and high-resolution images by resizing the image to match their resolution and cut-and-pasting the patches to generate augmented random samples.
Training a super-resolution network with CutBlur exhibits better performance than using CutMix or other regional dropout augmentations.

Even though DA methods like \cite{zhang2017mixup,yun2019cutmix,yoo2020rethinking} improve the performance of high- and low-level vision tasks, they are not adequate for image steganalysis.
A cover and its corresponding stego should be placed simultaneously in the same mini-batch  \cite{Boroumand2019, Zhang2020} to make the network focus on only the existence of stego signal regardless to image content.
Moreover, unlike image resizing, pixel modification in steganography is not spread uniformly but is adaptive to the image content to improve invisibility.
Therefore, DA for steganalysis must not only be augmented using the cover-stego pair but also be adaptive to steganography.

To meet the required conditions, we introduce BitMix, a DA for image steganalysis. 
BitMix is a DA with a regional dropout using cover-stego pairs that generate a soft target label that is adaptive to the steganography. 
BitMix generates training samples by mixing the cover-stego pairs by swapping random patches of the images, and it generates a target label for both the patch location and steganography signal inside the patch by measuring the ratio of the modified pixels in the swapped patch against those in the cover-stego image pair.
The target label of BitMix represents the confidence of the presence of a stego signal in the mixed area compared to the stego image in order to adapt to the steganography message embedding, whereas CutMix generates a fixed target label from the path size regardless of the patch location and steganography.
We present the detailed BitMix algorithm and explore the optimal parameters. 
The experimental results indicate that BitMix improves network performance in comparison to the existing DA methods.

\section{BitMix: Data Augmentation  for Steganalysis}
Let $C$, $S$ $\in \mathbb{R}^{W\times H}$ be a cover-stego image pair, where $y_C, y_S = 0, 1$ is the target label. BitMix mixes the cover ($C$) and stego ($S$) to generate a pair of new trainable images $(C_S, S_C)$ by swapping the patches in the same position between $C$ and $S$. We define the mixing operation as
\begin{equation}
\begin{aligned}
C_S &= \mathbf{M} \odot S + (\mathbf{1}-\mathbf{M}) \odot C  \\
S_C &= \mathbf{M} \odot C + (\mathbf{1}-\mathbf{M}) \odot S, 
\end{aligned}
\label{eq_bitmix}
\end{equation}
where $\mathbf{M}\in \{0,1\}^{W\times H} $ denotes a binary mask indicating where to swap one from another,  $\mathbf{1}$ is a binary mask filled with 1s, and $\odot$ is an element-wise multiplication.  $\mathbf{M}$ contains the bounding box $\mathbf{B}$ indicating the swapped regions in the image.
To sample the bounding box, we first set the maximum mix ratio to $\gamma<1$, which regulates the maximum bounding box size to $\gamma WH$. 
The coordinates of $\mathbf{B}=(r_x, r_y, r_w, r_h)$ are uniformly sampled with $\gamma$ according to:
\begin{equation}
    \begin{aligned}
        \gamma^\prime &\sim Unif(0, \gamma), \quad \gamma <1  \\
        r_w &=W\sqrt{\gamma^\prime}, \quad r_x \sim Unif(0, W-r_w) \\
        r_h &=H\sqrt{\gamma^\prime}, \quad r_y \sim Unif(0, H-r_h).
    \end{aligned}
    \label{eq_samplebox}
\end{equation}

With $\mathbf{M}$, we first calculate $\lambda$, which is the a ratio of the number of pixel modified in the swapped area to those in the cover-stego images, and we generate the target for $C_S$ and $S_C$ as
\begin{equation}
\begin{aligned}
\lambda &=\frac {\left\lVert\mathbf{M} \odot C - \mathbf{M} \odot S\right\lVert}  {\left\lVert C - S \right\lVert} \\
y_{C_S} &= \lambda, \quad y_{S_C} = 1-\lambda,
\label{BitMixRatio}
\end{aligned}
\end{equation}
which is similar to CutMix \cite{yun2019cutmix}, except that BitMix generates the target label to adapt to the steganography embedding rather than directly from the size of the bounding box.


\begin{figure}[t]
\centering
\subfigure[$S$ and $C-S$]{
  \centering
  \includegraphics[width=0.235\linewidth]{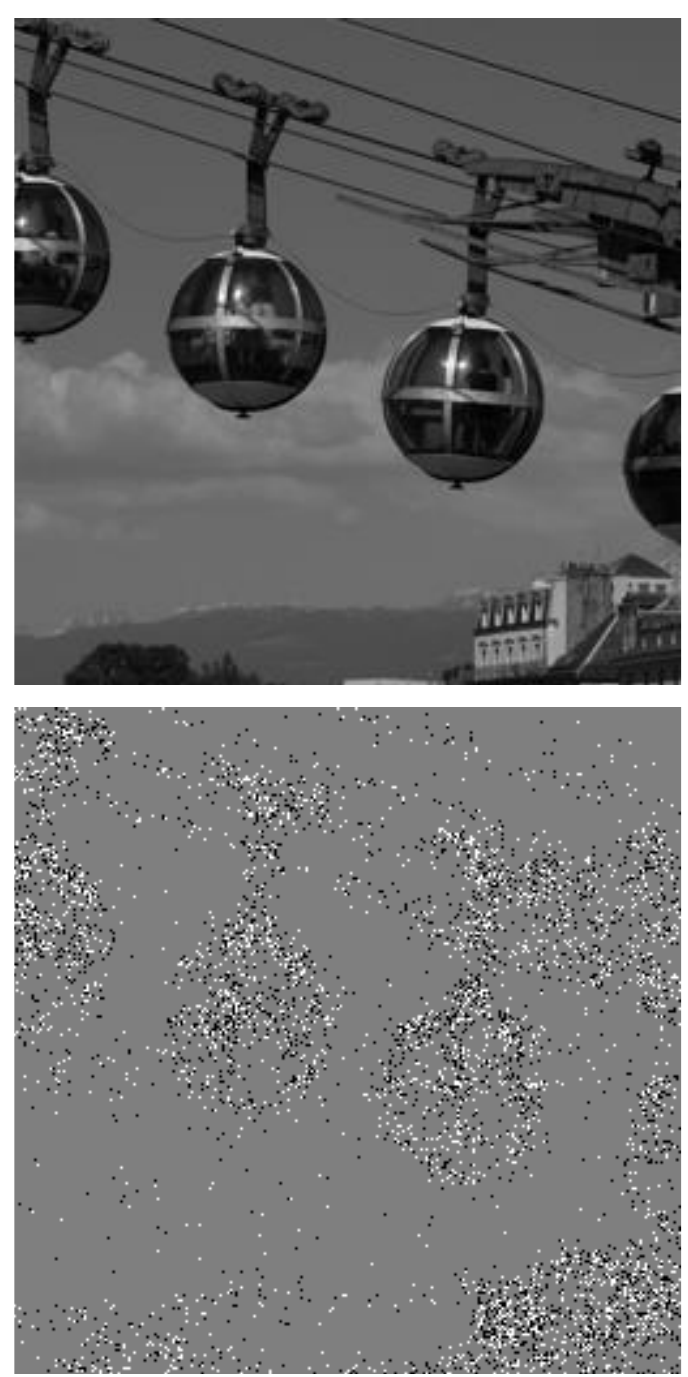} 
}
\hspace{-3mm}
\subfigure[$\mathbf{M}$]{
  \centering
  \includegraphics[width=0.235\linewidth]{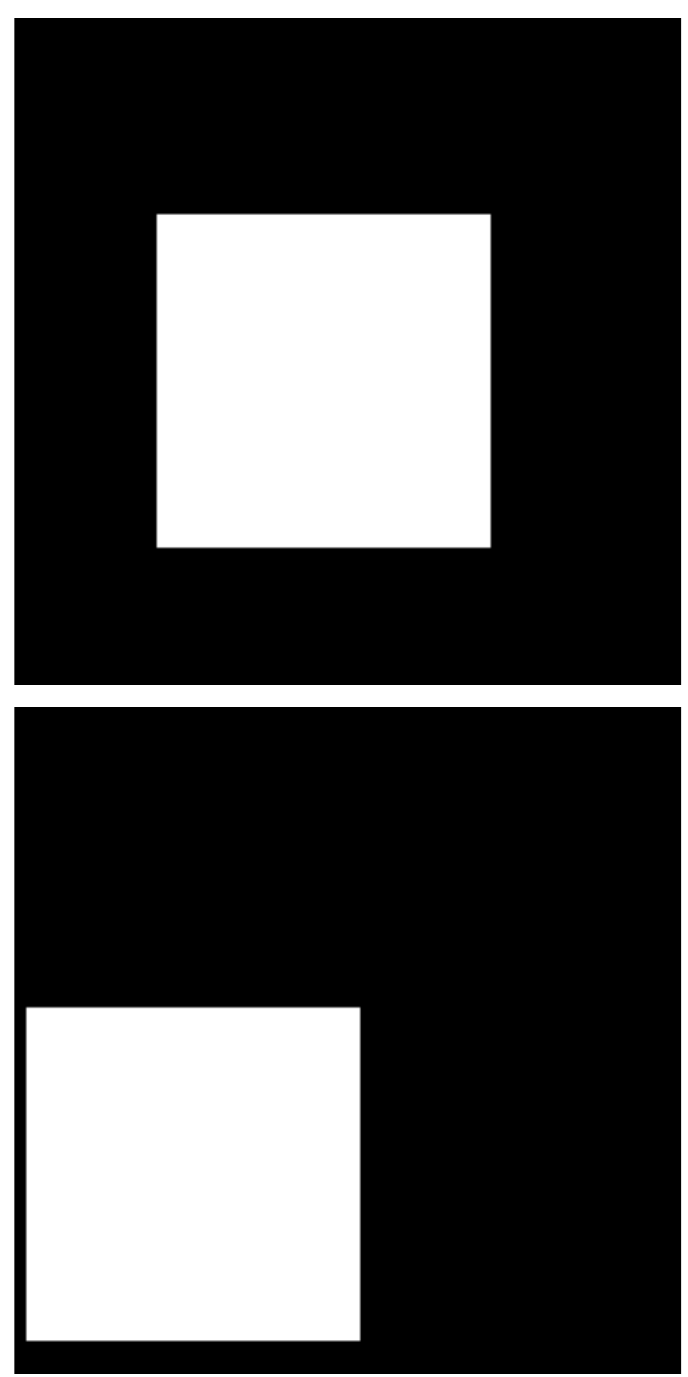} 
}
\hspace{-3mm}
\subfigure[$C_S$]{
  \centering
  \includegraphics[width=0.235\linewidth]{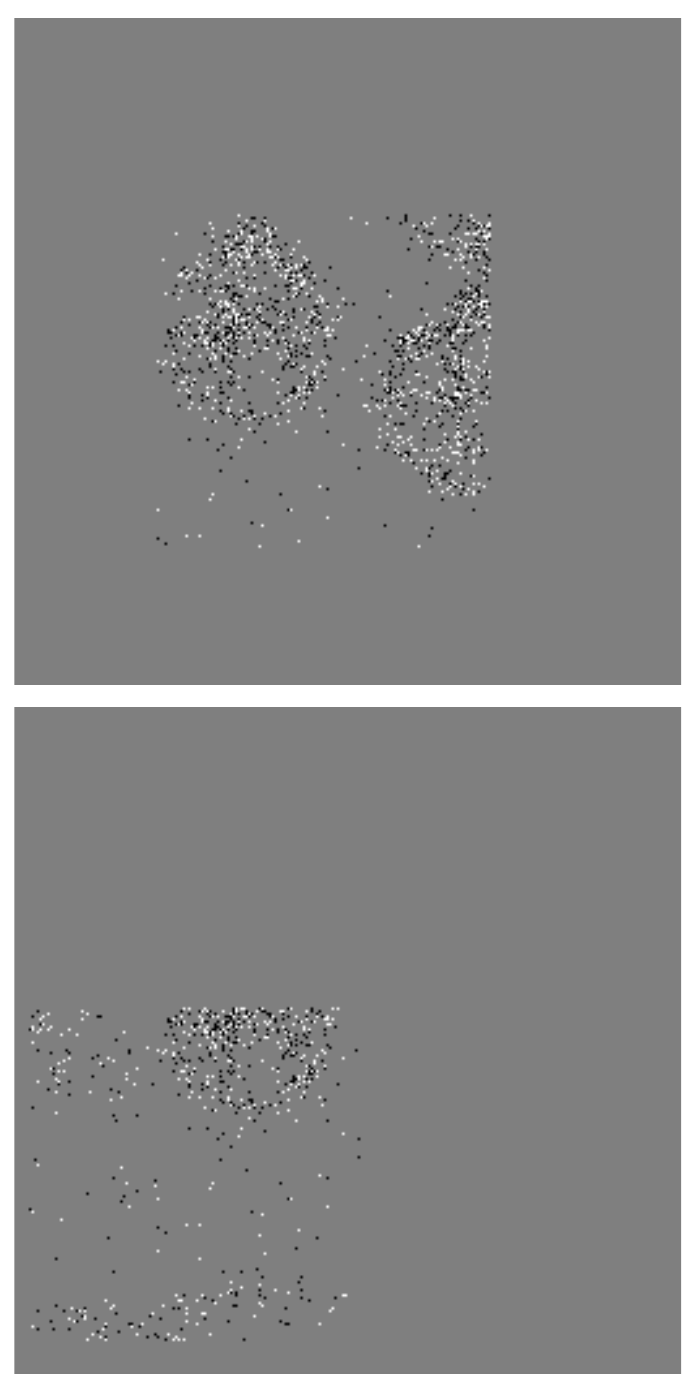} 
}
\hspace{-3mm}
\subfigure[$S_C$]{
  \centering
  \includegraphics[width=0.235\linewidth]{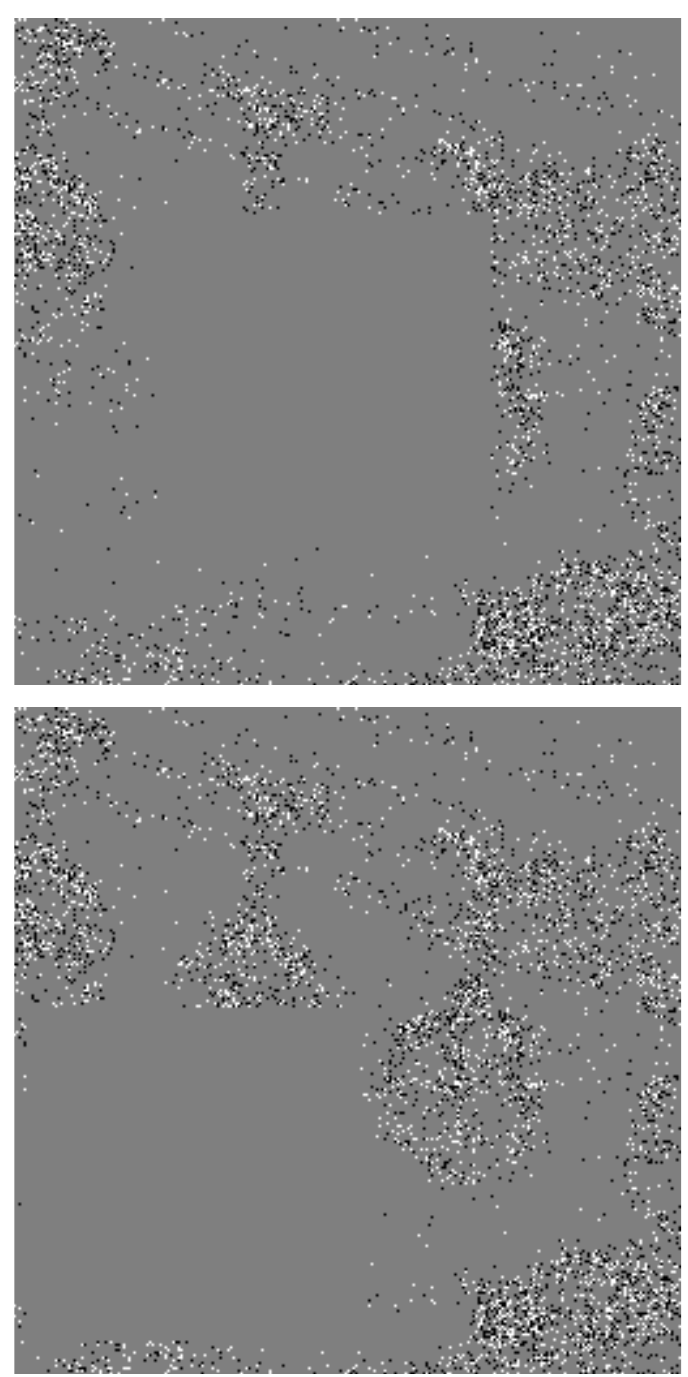} 
}
\caption{Overview of BitMix. The number of modified pixels is different with the position of the sampled bounding box.}\label{fig-algo}
\end{figure}

\begin{algorithm}[b]
\small
\caption{Applying BitMix on a singe mini-batch}
\textbf{Input} $\mathbf{C}, \mathbf{S}$: $N$ cover-stego pairs ($W \times H$), $\mathbf{y_C}$, $\mathbf{y_S}$: target labels\\
\textbf{Output} model input ($N$ normal and $N$ BitMix images), target labels\\
\textbf{Parameter} Maximum mix ratio $\gamma$ \\
\textbf{Initialization} $\mathbf{y_C}\leftarrow1$, $\mathbf{y_S}\leftarrow0$ 
\begin{algorithmic}[1]
\FOR{$i = 0$ to $N$}
\vspace{0.4mm}
    \STATE $\mathbf{C}[i]$, $\mathbf{S}[i]$ $\leftarrow$ randRotationFlip$(\mathbf{C}[i],\mathbf{S}[i])$
    \IF{$i<N/2$}  
    \vspace{0.4mm}
        \STATE $\gamma^\prime$ $\leftarrow$ Unif(0,$\gamma$) \COMMENT{Mask Sampling}
        \STATE $r_w$ $\leftarrow$ $ W\sqrt{\gamma^\prime}$, $r_h$ $\leftarrow$ $  H\sqrt{\gamma^\prime}$
        \STATE $r_x$ $\leftarrow$ Unif$(0,W-r_w$), $r_x$ $\leftarrow$ Unif$(0,H-r_h$) 
        \STATE $M$ $\leftarrow$ binaryMask$(W,H,r_x,r_y,r_w,r_h)$
        \STATE $\mathbf{C}[i]\leftarrow M \odot S + (1-M) \odot C$ \COMMENT{Mix Images}
        \STATE $\mathbf{S}[i]\leftarrow M \odot C + (1-M) \odot S$
        \STATE $\lambda \leftarrow\left\lVert M \odot C - M \odot S\right\lVert /  \left\lVert C - S  \right\lVert$ \STATE $\mathbf{y_C}[i]\leftarrow\lambda, \mathbf{y_S}[i]\leftarrow1-\lambda$ \COMMENT
        {Embedding adaptive targets}
    \ENDIF
\ENDFOR
\STATE model input $\leftarrow$ $[\mathbf{C}, \mathbf{S}]$, target labels $\leftarrow$ $[\mathbf{y_C}, \mathbf{y_S}]$
\end{algorithmic}
\label{algo_bitmix}
\end{algorithm}

To make the network focus on the existence of a stego signal, each cover-stego pair in a mini-batch is augmented with a random rotation and flip \cite{yedroudj2018yedroudj,Boroumand2019, Zhang2020}. 
We observed that simply applying BitMix to all image pairs cause the validation loss to diverge.
Therefore, we apply BitMix to half of the cover-stego pairs for each mini-batch. 
Each input-target pair is fed to the network to minimize the binary cross-entropy loss.
The overall code-level details are presented in Algorithm \ref{algo_bitmix}. 

\begin{figure}[t]
\centering
\includegraphics[width=1.0\linewidth]{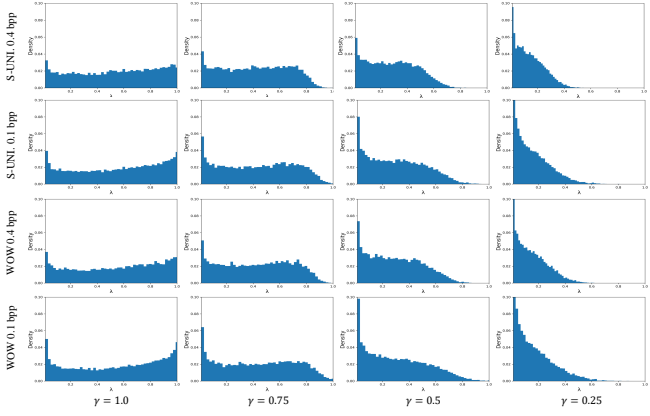}
\caption{Distribution of the modified pixel swapped ratio $\lambda$ for each steganography method and bits per pixel (bpp) with respect to the maximum mix ratio $\gamma$.}\label{fig-dist}
\end{figure}

\vspace{-2mm}

\section{Discussions}\label{sec-discuss}
Fig.~\ref{fig-algo} (a) is an example of a stego image with a resolution of $256\times256$, for which the number of pixels modified from the steganography is 5,202.
Fig.~\ref{fig-algo} (c), (d) illustrates the results of patch-swapping using a different bounding box with an equal size to the resolution (25\%), where the numbers of changed pixels are 621 and 1,416. 
With BitMix, the target labels for each case are  $(0.12, 0.88)$ and $(0.27, 0.73)$, which is same as $(0.25, 0.75)$ if we use CutMix \cite{yun2019cutmix}, which does not consider the steganography. 
The target label generated using BitMix is adaptive to the steganography embedding, as it refers to the ratio of the actual pixel modification in the area where the pixel might have been modified.

We observe that training with a different maximum mix ratio $\gamma$ produces a different performance. 
Before seeking the optimal value for $\gamma$, we first calculate the distribution of the modified pixel swapped ratio $\lambda$, as the maximum mix ratio $\gamma$ changes to 1, 0.75, 0.5 and 0.25 for the two spatial steganography at 0.4, 0.1 bits per pixel (bpp) (see Fig. \ref{fig-dist}). 
The distribution is concentrated down to $0$ as $\gamma$ decreases, whereas the distribution follows to $Unif(0,\gamma)$ when using CutMix \cite{yun2019cutmix}.
Moreover, if $\gamma$ is fixed, the distribution of $\lambda$ does not change much even if the steganograpic method or bpp changes.
With this observation, we determined the value of $\gamma$ that provides the best result for one training case and applied it to the remaining experiments rather than searching for the optimal $\gamma$ value for each steganograpic method, bpp, and model.

\section{Training Details}
We used the union of BOSSBase 1.01 and BOWS2, with each grayscale image resized to $256 \times 256$ using the MATLAB default resizing function. The entire BOWS2 dataset was used for training, and we randomly divided the images from BOSSBase into training, validation, and test sets with a ratio of 4:1:5. We evaluated BitMix using two spatial domain steganographic method, WOW and S-UNIWARD at 0.4, 0.3, 0.2 and 0.1 bpp.

We used two state-of-the-art CNN-based steganalysis, SRNet \cite{Boroumand2019} and ZhuNet \cite{Zhang2020}, as our baseline models. 
Each network model was trained with a paired mini-batch that consists of a cover and its corresponding stego images placed in a single mini-batch (16 cover-stego pairs). 
Each cover-stego pair was first augmented using the random rotation in multiples of 90$\degree$ and random flipping, and BitMix was applied to half of the cover-stego pairs in the mini-batch. 
All models were trained for up to 200 epochs using an AdamW optimizer with an initial $lr=10^{-4}$, which reduced to $2.5\times10^{-5}$ and $10^{-5}$ after 75 and 150 epochs, respectively.
We set the weight decay as $0.1\times lr$ when changing the learning rate manually.

We first trained the network for each steganograpic method at 0.4 bpp with these hyperparameters. 
For the remaining payloads at 0.3, 0.2, and 0.1 bpp, we trained the model via curriculum learning in the same manner as in \cite{Boroumand2019}.
These models were fine tuned with the best validation error model at 0.4, 0.3, and 0.2 bpp, respectively, with an initial  $lr=2.5\times10^{-5}$, which was reduced to $1\times10^{-5}$ after 50 epochs.
Each trained model was measured using the error rate $P_E$ and area under the curve (AUC) where $P_E=min_{P_{FA}} \frac{1}{2} (P_{FA} + P_{MD})$, and $P_{FA}$ and $P_{MD}$ are the false-alarm and missed-detection possibilities.

\begin{table}[t]
\caption{$P_E$ and AUC of SRNet trained with BitMix with a different $\gamma$.}
\scriptsize
\centering
\begin{tabu} to \linewidth{X[c,0.8]|X[c,0.8] X[c,0.8] X[c,0.8] X[c,0.8] X[c,1.0] X[c,1.6]}  
\hline\hline
\multirow{2}{*}{Metric} & \multicolumn{6}{c}{$\gamma$}\\ 
\cline{2-7}
& 1     & $0.75$  & $0.5$   & $0.25$ & $0.0625$ & 0 (baseline) \\ \hline
$P_E$ & 0.1161 & 0.1159 & 0.1124 & \textbf{0.1101} & 0.1117 & 0.1160      \\
AUC & 0.9620 & 0.9645 & 0.9647 & \textbf{0.9657} & 0.9619 & 0.9640 \\
\hline\hline
\end{tabu}
\label{table_gamma}
\end{table}

\begin{figure}[t]
\centering
\subfigure[$S_C$ with $\lambda=0.9$]{
  \centering
  \includegraphics[width=0.47\linewidth]{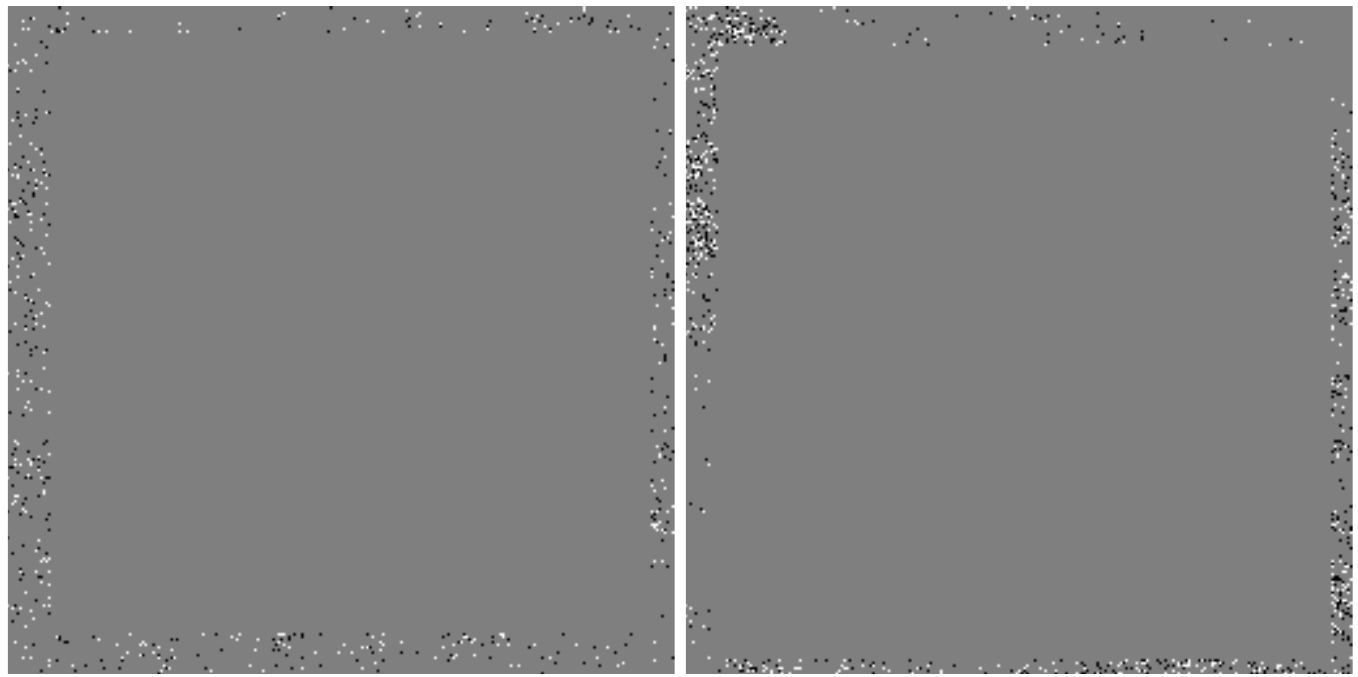}
}
\hspace{-3mm}
\subfigure[$C_S$ with $\lambda=0.1$]{
  \centering
  \includegraphics[width=0.47\linewidth]{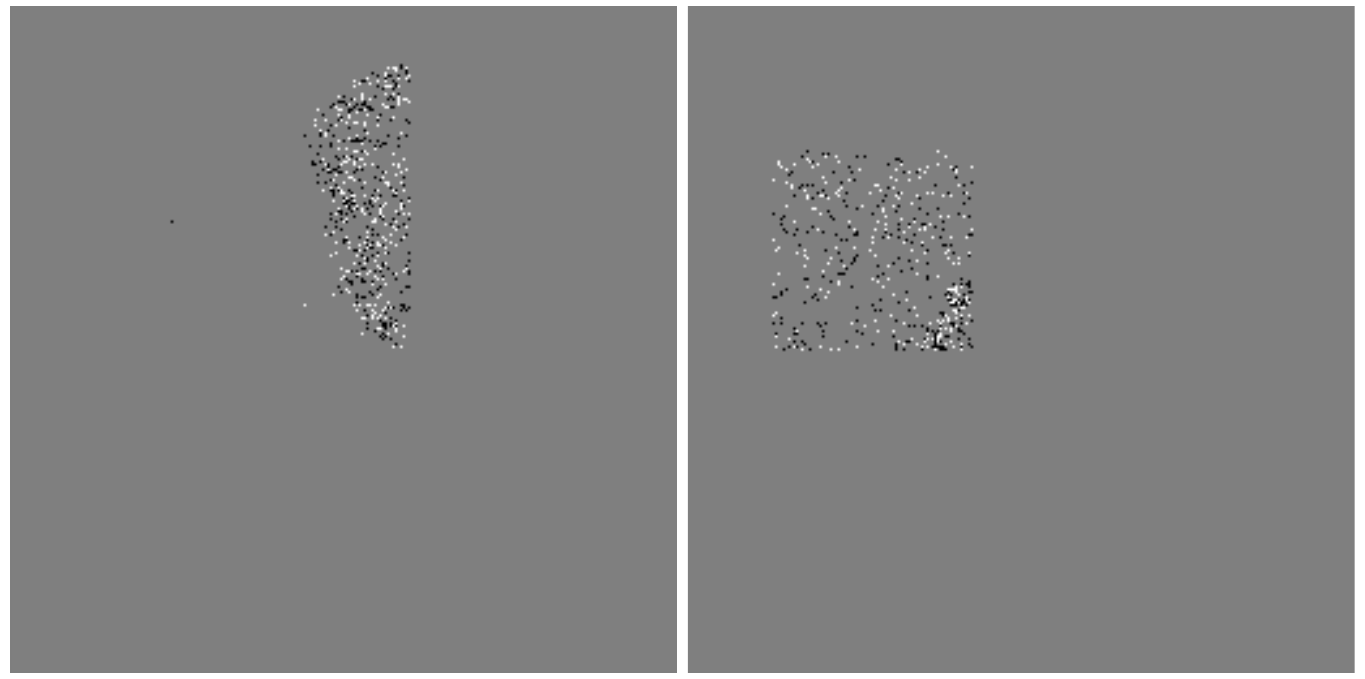}
}
\caption{Location of modified pixels in the images augmented with different maximum mix ratio $\gamma$. Each image has the same target label, ($0.1$), but is made from (a) large and (b) small $\gamma$.}\label{fig-box}
\end{figure}

\begin{figure*}[t]
\centering
\includegraphics[ width=\linewidth]{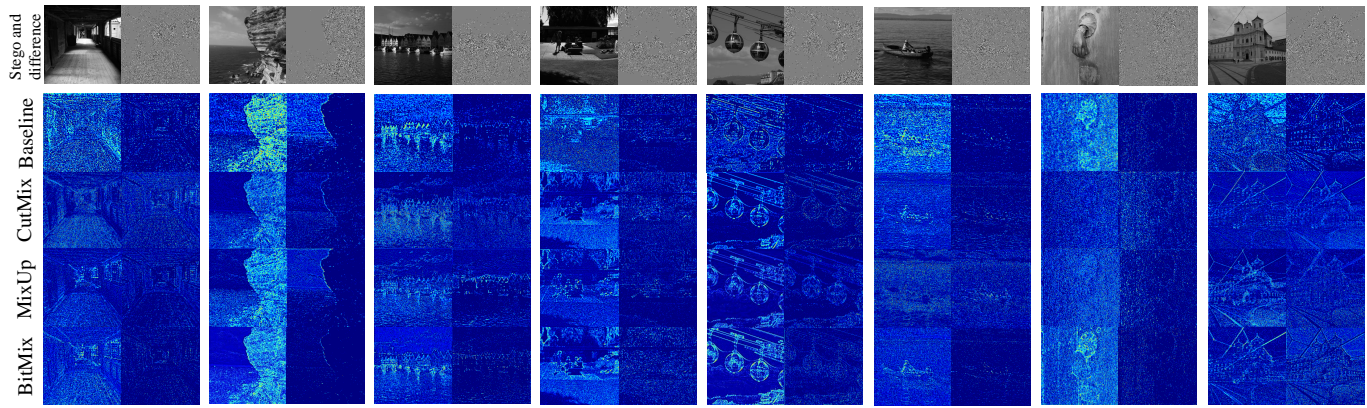}
\caption{Feature discrimination ability of DA methods.  First row: stego and location of pixel modification.  Second to fifth rows: Grad-CAM visualization of last unpooled layer of SRNet trained on S-UNIWARD at 0.4 bpp with different DA methods. Each sub-column represents a Grad-CAM stego-image for each stego (left) and cover (right) class.}\label{fig-cam}
\end{figure*}

\section{Results}
We trained SRNet on S-UNIWARD at 0.4 bpp using a different maximum mix ratio $\gamma$ (see Table~\ref{table_gamma}). 
When $\gamma=0$, it is equal to the baseline; therefore, selecting an appropriate $\gamma$ is important.
When $\gamma=1$, the detection error rate and AUC are slightly worse than at the baseline, but the performance improves as $\gamma$ decreases from 0.75 to 0.25, and it deteriorates as $\gamma$ decrease to zero again ($0.0625$).

The reason the performance improves when $\gamma$ decreases is the distribution of the modified pixels in the image with respect to $\gamma$ and $\lambda$ (see Fig.~\ref{fig-box}).
If $\lambda=0.9$, $S_C$ contains 10\% of the modified pixels in the original stego with a target label of 0.9.
When the area of the bounding box is close to the image resolution, the modified pixels are placed at the outer parts, which makes the pixel distribution static even if the image changes.
Meanwhile, when $\lambda=0.1$, $C_S$ contains 10\% of the modified pixels from the stego with a target label of 0.9, and the modified pixels are randomly distributed for different images.
Even if both cases have the same target label, the distribution of the modified pixels in the results with a small $\gamma$ is much more diverse than using a large $\gamma$.
Based on the experiment results and this analysis, the rest of the experiments are conducted with $\gamma=0.25$, which provides the best performance. 


Table~\ref{tab_mainexp} reports the performance of SRNet and ZhuNet with and without BitMix for each spatial image steganography at 0.4 to 0.1 bpp. 
For S-UNIWARD at 0.4 bpp, SRNet with BitMix demonstrates an 11.01\% detection error rate, which is lower than the baseline at 11.6\%, and ZhuNet with BitMix achieves a 14.7\% detection error rate, which is lower than the baseline at 16.0\%. 
For 0.3 bpp to 0.1 bpp which are fine-tuned using the best model for 0.4 bpp to 0.2 bpp, a better detection error rate is produced than the baseline for both SRNet and ZhuNet. 
Except for the SRNet trained on S-UNIWARD at 0.2 bpp, every trained model exhibits a better AUC than the baseline. 
For WOW, SRNet exhibits a higher performance than that of the baseline, whereas ZhuNet exhibits a lower error rate for all bpp but a lower AUC at 0.2 and 0.4 bpp.

 

To verify the effectiveness BitMix, we compared its performance with other DA methods.
We trained SRNet and ZhuNet on S-UNIWARD at 0.4 bpp using CutMix \cite{yun2019cutmix} and MixUp \cite{zhang2017mixup}.
For a fair comparison, we applied each augmentation to half of the cover-stego pairs in each mini-batch and set the maximum size of the sampled bounding box for CutMix to 0.25, which is the same as in BitMix.
BitMix has a better detection error rate and AUC than the baseline and other DA methods for both SRNet and ZhuNet (see Table~\ref{table_aug}).
Although other DA methods exhibits better error rate than the baseline, CutMix had a lower AUC for SRNet, and MixUp had a lower AUC for both networks. 

In addition, we examined which part of the image the network reacts to when training using the DA method.
To check this, we used gradient class activation map (Grad-CAM), which can be applied to the intermediate layer (see Fig.~\ref{fig-cam}). We visualized Grad-CAM in the last unpooled layer of SRNet for stego images for the stego and cover classes trained with each DA method. 
We only considered SRNet because ZhuNet employs a fixed kernel initialization and early pooling, whereas SRNet is a full end-to-end model with multiple unpooled layer, which allows for a clear interpretation of the low-level features.

For the stego class (left side of each column), the highlighted areas refer to the area that contributes to the stego class. 
The Grad-CAM of BitMix has the highest contrast and most precise location of the modified pixels among the DA methods. 
The network trained with BitMix has a better discrimination ability regarding the area where the steganography embedding modifies the pixels regardless of the image contents (sixth and seventh columns). 
For the cover class (right side), the Grad-CAM of BitMix shows has contrast and contains less information about the image content than other DA methods (third and fifth columns).
We conclude that BitMix with optimized parameters improves performance by guiding the network to the presence of the steganography signal more effectively than the other aforementioned DA methods.

\begin{table}[t]
\caption{$P_E$ and AUC for spatial steganalysis networks using BitMix}
\scriptsize
\centering
\begin{tabu} to \linewidth{X[c,0.5]|X[c,0.6] |X[c,1.0] X[c,1.0] X[c,1.0] X[c,1.0]} 
\hline\hline
\multirow{2}{*}{Method} &\multirow{2}{*}{Model} &\multicolumn{4}{c}{bpp} \\
\cline{3-6} 
 & & 0.1 & 0.2 & 0.3 & 0.4 \\ \hline
\multirow{4}{*}{S-UNI} & SRNet         & 0.3122/0.7711     & 0.2116/\textbf{0.8891}    & 0.1506/0.9427     & 0.1160/0.9640    \\
                           & +BitMix  & \textbf{0.3073/0.7834}     & \textbf{0.2101}/0.8843    & \textbf{0.1453/0.9454}     & \textbf{0.1101/0.9657}   \\\cline{2-6} 
                           & ZhuNet        &    0.3578/0.7182 & 0.2633/0.8388    & 0.1947/0.9059     & 0.1600/0.9359    \\
                           & +BitMix & \textbf{0.3501/0.7345}    & \textbf{0.2529/0.8501}   & \textbf{0.1941/0.9129}     & \textbf{0.1470/0.9441}    \\ \hline
\multirow{4}{*}{WOW}       & SRNet         & 0.2664/0.8309     & 0.1792/0.9220    & 0.1332/0.9531     & 0.0973/0.9733    \\
                           & +BitMix  & \textbf{0.2622/0.8347}     & \textbf{0.1723/0.9226}    & \textbf{0.1266/0.9561}     & \textbf{0.0945/0.9737}    \\\cline{2-6}
                           & ZhuNet        &   0.3001/0.7977  & 0.2053/\textbf{0.8990}    & 0.1557/0.9399     & 0.1143/\textbf{0.9651} \\
                           & +BitMix & \textbf{0.2968/0.8005}      &\textbf{0.2031}/0.8963    &    \textbf{0.1519/0.9401}  &\textbf{0.1122}/0.9646   \\\hline\hline
\end{tabu}
\label{tab_mainexp}
\end{table}

\begin{table}[t]
\caption{$P_E$ and AUC of SRNet using DA on S-UNIWARD at 0.4 bpp.}
\scriptsize
\centering
\begin{tabu} to \linewidth{X[c,0.8]|X[c,0.9]|X[c,0.8] X[c,1.5] X[c,1.4] X[c,0.8]}  
\hline\hline
Model & Metric & Baseline & CutMix \cite{yun2019cutmix}& MixUp \cite{zhang2017mixup}& BitMix \\ \hline
\multirow{2}{*}{SRNet} & $P_E$   & 0.1160     & 0.1143  &  0.1148     & \textbf{0.1101}  \\
& AUC  & 0.9640       & 0.9635       & 0.9626      & \textbf{0.9657}   \\ \hline
\multirow{2}{*}{ZhuNet} & $P_E$   & 0.1600     & 0.1506  &  0.1567     & \textbf{0.1470}  \\
& AUC  & 0.9359       & 0.9423       & 0.9354      & \textbf{0.9441}   \\

\hline\hline
\end{tabu}
\label{table_aug}
\end{table}

\section{Conclusion}
Steganalysis is required to detect very subtle noise in stego images, which is different from high-level vision tasks.
State-of-the-art CNN-based steganalysis is benefited from DA but it produces a fixed number of images that limits the effects of DA. 
Our goal is to design a DA method for steganalysis that both preserves the steganography signal and increases the number of different generated samples by embedding adaptive target labels.
To achieve this purpose, we proposed BitMix, a DA for spatial image steganalysis.
BitMix generates augmented samples by swapping patches from cover-stego image pairs and generates their target labels to adapt to the location of the modified pixels from steganography embedding.
We explored the optimal maximum mix ratio to generate dynamically modified pixel distributions.
Moreover, we found that BitMix applies to various situations based on the observation of  a fixed label distribution for different steganographic methods at different bits per pixel with a fixed maximum mix ratio.
Compared to the performance of fixed augmentation, the network trained using BitMix exhibits better performance in most cases.
Furthermore, BitMix improves the performance and discrimination ability of the stego signal than when simply applying the existing DA methods.


\vspace{+2mm}
\noindent I.-J. Yu, W. Ahn, S.-H. Nam, and H.-K. Lee (\textit{School of Computing, Korea Advanced Institute of Science and Technology, Daejeon, Republic of Korea})

\noindent E-mail: ijyu@mmc.kaist.ac.kr

\bibliographystyle{iet}

\end{document}